\documentstyle[prc,preprint,graphicx,floats,aps]{revtex}

\renewcommand\topfraction{.85}
\renewcommand\textfraction{.15}

\begin{document}
 \newcommand {\nucleus}[2] {\mbox{${}^{#2}{#1}$}}

\tighten
\draft
\title{The role of $\nu$-induced reactions on lead and iron in neutrino
       detectors}

\author{E.Kolbe$^{1,}$\footnote{Permanent address: 
Departement f\"ur Physik und Astronomie der Universit\"at Basel,
Basel, Switzerland} and
K. Langanke$^2$}
\address{
$^1$ Institut f\"ur Kernphysik I, Forschungszentrum Karlsruhe, Germany \\
$^2$Institute for Physics and Astronomy, University of Aarhus, Aarhus, Denmark 
}
\date{\today}
\maketitle


\begin{abstract}
We have calculated cross sections and branching ratios for neutrino 
induced reactions on $^{208}$Pb and $^{56}$Fe for various supernova and
accelerator-relevant neutrino spectra. 
This was motivated by the facts that lead and iron will be used on one hand 
as target materials in future neutrino detectors, on the other hand have 
been and are still used as shielding materials 
in accelerator-based experiments.
In particular we study the inclusive 
$^{56}$Fe($\nu_e,e^-$)$^{56}$Co and $^{208}$Pb($\nu_e,e^-$)$^{208}$Bi 
cross sections and calculate the neutron energy spectra following
the  decay of the daughter nuclei.
These reactions give a potential background signal in the 
KARMEN and LSND experiment and are discussed as a detection scheme for
supernova neutrinos in the proposed OMNIS and LAND detectors. We also study the
neutron-emission following the neutrino-induced neutral-current
excitation of $^{56}$Fe and $^{208}$Pb.
\end{abstract}

\pagestyle{plain}
\newpage
\section{Introduction}
Neutrinos play a decisive role in many aspects of astrophysics and
determining their properties is considered the most promising gateway to
novel physics beyond the standard model of elementary particle physics. 
Thus detecting and studying accelerator-made or astrophysical neutrinos is a
forefront research issue worldwide with many ongoing and planned
activities.

One of the fundamental questions currently investigated is whether
neutrinos have a finite mass. This question can be answered by the
potential detection of neutrino oscillations which would establish the 
existence of at least one family of massive neutrinos. 
Furthermore, the existence of massive neutrinos might have profound 
consequences on many branches of cosmology and astrophysics, 
e.g. the expansion of the 
universe and the formation of galaxies, while neutrino oscillations can
have interesting effects on supernova nucleosynthesis \cite{Raffelt}.

From the many experiments directly searching for neutrino oscillations,
only the LSND collaboration has reported positive candidate events 
\cite{At95}. 
Indirect evidence for neutrino oscillations arises from the deficit 
of solar neutrinos, as observed by all solar-neutrino detectors
\cite{Bahcall89}, and the suppression and 
its angular dependence of events induced by atmospheric $\nu_\mu$ neutrinos
in Superkamiokande
\cite{Fu98,St96}.
Due to the obvious importance, the oscillation results implied from 
these experiments will be cross-checked by future long-baseline 
experiments like MINOS \cite{Minos}. From the detectors currently operable
KARMEN has a neutrino-oscillation sensitivity similar 
to the LSND experiment. Currently, the KARMEN collaboration does not observe 
oscillations covering most of the oscillation parameter space for 
the positive LSND result
\cite{Ar98}.

A type II supernova releases most of its energy in terms of neutrinos.
Supernova neutrinos from SN87a
had been observed by the Kamiokande and IMB detectors \cite{Kamio,IMB}
and have confirmed the general supernova picture.
The observed
events were most likely  due to ${\bar \nu_e}$ antineutrinos. However,
the models predict distinct differences in the 
neutrino distributions for the various families and thus a more
restrictive test of the current supernova theory requires the abilities
of neutrino spectroscopy by the neutrino detectors.
Current (e.g. Superkamiokande and SNO)
and future detectors (including the proposed OMNIS \cite{Omnis,Cline}
and LAND \cite{Land}
projects)
have this  capability 
and will be able to distinguish between the different neutrino types
and determine their individual spectra. 
For the water \v{C}erenkov detectors
(SNO and Superkamiokande) $\nu_x$ neutrinos can be detected by specific
neutral-current events \cite{SNO,Langanke96}, 
while the OMNIS and LAND detectors are
proposed to detect neutrons spallated from target nuclei by charged- and
neutral-current neutrino interactions. 

Some of the supernova-neutrino or neutrino-oscillation  
detectors use iron or lead as detector material
(e.g. MINOS, LAND  and OMNIS) or have adopted steel (LSND, KARMEN) 
and lead (LSND) shielding. 
Thus, precise theoretical estimates of the neutrino-induced
cross sections on Fe and Pb are required for a reliable 
knowledge of the detection signal or the appropriate 
simulation of background events.
We note that the KARMEN collaboration has recently used its  
sensitivity to the $^{56}$Fe($\nu_e,e^-$)$^{56}$Co background events 
to determine 
a cross section for this reaction \cite{Maschuw}. 
In Ref. \cite{Kolbe99} we have calculated this cross section in a hybrid
model in which the allowed transitions have been studied based on the
interacting shell model, while the forbidden transitions were calculated
within the continuum random phase approximation. In this paper we extend
this investigation and study the charged- and neutral current reactions
on $^{56}$Fe and $^{208}$Pb for various accelerator-based and supernova
neutrino distributions. In particular, we determine 
the $^{208}$Pb($\nu_e,e^-$) 
cross sections for the LSND neutrino spectra which will serve for even
improved
background simulations for this detector. Our calculations of 
supernova neutrino
reaction cross sections on $^{56}$Fe and $^{208}$Pb  are aimed to
guide the design of supernova neutrino detectors like 
OMNIS and LAND. With this
goal in mind we have calculated the energy spectrum of neutrons
knocked-out by the charged-current or neutral-current neutrino-induced
excitation of $^{56}$Fe and $^{208}$Pb. To allow also the exploration of
potential oscillation scenarios we have calculated the cross sections
and neutron spectra for various supernova neutrino spectra.

\section{Theoretical model}

Besides the total cross sections, the partial cross sections for 
neutrino-induced particle knock-out are of relevance to estimate 
the signal and background of the various detectors. 
We will calculate these partial cross sections in a two-step 
process (e.g. for the charged-current reaction):

\begin{displaymath} 
\! \! \! \begin{array}{rcl} 
   \underbrace{ \begin{array}{rcl}
      \nu + {}_{Z}X_{N} & \rightarrow & l + {}_{Z+1}X^*_{N-1} 
   \end{array} }_{\rm 1. \; \;  \; \; RPA} 
   & \Longrightarrow &
   \underbrace{ \begin{array}{rcl}
      {}_{Z+1}X^*_{N-1} & \rightarrow & \left\{ \begin{array}{rcl} 
                            {}_{Z+1}X_{N-2}  &+&{\rm n} \\ 
                            {}_{Z  }X_{N-1} &+&{\rm p} \\ 
                            {}_{Z-1}X_{N-3} &+&\alpha \\
                            {}_{Z+1}X_{N-1} &+&\gamma 
                         \end{array} \right. 
   \end{array} }_{\rm 2. \; \; Statistical \; \; Model}
\end{array} 
\end{displaymath}

In the first step, we calculate
the $\nu$-induced  spectrum 
$\frac{d\sigma}{d\omega} (\omega)$
in the daughter nucleus 
at excitation energy $\omega$. We consider multipole excitations of
both parities and
angular momenta $\lambda \leq 9$, using the formalism developed
in\cite{Walecka}. These multipole operators, denoted by $\lambda^\pi$,
depend on the momentum transfer $q$.

Our strategy to calculate 
$\frac{d\sigma}{d\omega} (\omega)$ has been different for $^{56}$Fe and
$^{208}$Pb. For $^{56}$Fe we adopt the same hybrid model which has
already been successfully applied in \cite{Kolbe99}. That is, we
calculate all nuclear responses within the random phase approximation
(RPA). However, the RPA does not usually recover sufficient nucleon-nucleon
correlations to reliably reproduce the quenching and fragmentation of
the Gamow-Teller (GT) strength distribution in nuclei.
For this reason we determine the response of the $\lambda^\pi = 1^+$
operator on the basis of an interacting shell model calculation
performed within the complete $pf$ shell. Such a study has been proven
to reproduce the experimental GT$_-$ 
(in which a neutron is changed into a proton) 
and GT$_+$
(in which a proton is changed into a neutron) 
distributions on $^{56}$Fe 
well \cite{Caurier99}, 
if the response is
quenched by a universal factor $(0.74)^2$ \cite{Brown88,Langanke95,Martinez96}.
However,  the GT operator
corresponds to the appropriate $\lambda^\pi=1^+$ operator only in the
limit of momentum transfer $q \rightarrow 0$. As it has been pointed
out in \cite{Kolbe99,Hektor}, the consideration of the finite-momentum
transfer in the operator results in a reduction of the cross sections,
caused by the destructive interference with the higher-order operator
$\tau {\vec \sigma} {\vec r} \cdot {\vec p}$. To account for the effect
of the finite momentum transfer we have performed RPA calculations for
the $\lambda^\pi =1^+$ multipole operator at finite momentum transfer
$q$ (i.e. $\lambda(q)$) and for $q=0$ (i.e. $\lambda(q=0)$) and have
scaled the shell model GT strength distribution by the ratio of
$\lambda(q)$ and $\lambda(q=0)$ RPA cross sections. 
The correction is rather small for $\nu_e$ neutrinos stemming
from muon-decay-at-rest (i.e. for LSND and KARMEN) 
or for supernova $\nu_e$ neutrinos. The correction is, however, sizeable
if neutrino oscillations occur in the accelerator-based experiments or a
supernova \cite{Hektor}.

For $^{208}$Pb a converged shell-model calculation of the GT strength
distribution is yet not computationally feasible. Thus we have also
calculated the $\lambda^\pi =1^+$ response within the RPA approach. Note
that our RPA approach fulfills the Fermi and Ikeda sumrules.
As the $S_{\beta^+}$
strength (in this direction a proton is changed into a neutron)
is strongly suppressed for $^{208}$Pb, the Ikeda
sumrule fixes the $S_{\beta^-}$ strength. 
We have renormalized the $\lambda^\pi = 1^+$ strength in $^{208}$Pb by
the universal quenching factor which, due to a very slight
$A$-dependence is recommended to be $(0.7)^2$ in $^{208}$Pb
\cite{Martinez96}.
Thus the Ikeda sumrule reads
$S_{\beta_-} - S_{\beta_+} \approx S_{\beta_-} = 3 \cdot (0.7)^2 \cdot
(N-Z)$.
For the other multipole operators  no experimental evidence exists for
such a rescaling and we have used the RPA response.

In our RPA calculations we have chosen the
single-particle energies  from an appropriate Woods-Saxon 
potential, which has been adjusted to 
reproduce the relevant particle thresholds.
As residual interaction we used the zero-range Landau-Migdal force from
\cite{Rinker}. However, it is well known that this parameterization places
the isobaric analog state in $^{208}$Bi at too high an energy. This is
cured by changing the parameter, which multiplies the 
$\tau_i
\cdot \tau_j$ term in the interaction from 
$f_0^\prime =1.5$ to the value 0.9
\cite{Plumley}.  
After this adjustment the IAS is very close ($E_{IAS}=15.4$ MeV)
to the experimental position (15.16 MeV). Furthermore, our RPA
parametrization has been demonstrated to describe the $^{208}$Pb(p,n)
reaction data at small forward angle well \cite{Plumley}. 
Our RPA approaches are described in details
in Refs. \cite{Kolbe92,Kolbe99a}. We note that this approach gives quite
satisfying results for neutrino scattering \cite{Kolbe92,Kolbe94,Kolbe95}, 
muon capture \cite{Kolbe94a} and electron scattering \cite{Kolbe96}. 

After having determined the neutrino-induced excitation spectrum in the 
daughter nucleus, we calculate in
the second step for each final state 
with well-defined energy, angular momentum, and parity
the branching ratios into the various decay channels
using the statistical model code SMOKER \cite{Co91}.
The decay channels considered are 
proton, neutron, $\alpha$, and $\gamma$ emission. 
As possible
final states in the residual nucleus the SMOKER code considers
the experimentally known levels supplemented at higher energies
by an appropriate level density formula. 
Note, that the SMOKER code has been successfully applied 
to many astrophysical problems and that we empirically found 
good agreement between p/n branching-ratios calculated with
SMOKER and within continuum RPA 
for several neutral current reactions on light nuclei \cite{Kolbe92}.

As supernova and accelerator-produced neutrinos have an energy spectrum,
the final results (total and partial cross sections) are obtained by
folding with the appropriate neutrino spectra.

\section{Results}

\subsection{Reactions induced by decay-at-rest neutrinos}
  
The $\nu_e$ neutrinos produced in the muon decay-at-rest (DAR), have the
characteristic Michel energy spectrum
\begin{equation}
n(E_\nu) = \frac{96 E_\nu^2}{M_\mu^4} (M_\mu - 2 E_\nu),
\end{equation}
where $M_\mu$ is the muon mass and $E_\nu$ the neutrino energy.

Our calculated excitation spectrum for the
$^{56}$Fe($\nu_e,e^-$)$^{56}$Co reaction is shown in \cite{Kolbe99}.
Fig.~1 shows the RPA response for the $^{208}$Pb($\nu_e,e^-$)$^{208}$Bi 
reaction, calculated for a muon decay-at-rest neutrino spectrum. 
\begin{figure}[thb]
     \includegraphics[angle=90,height=9cm]{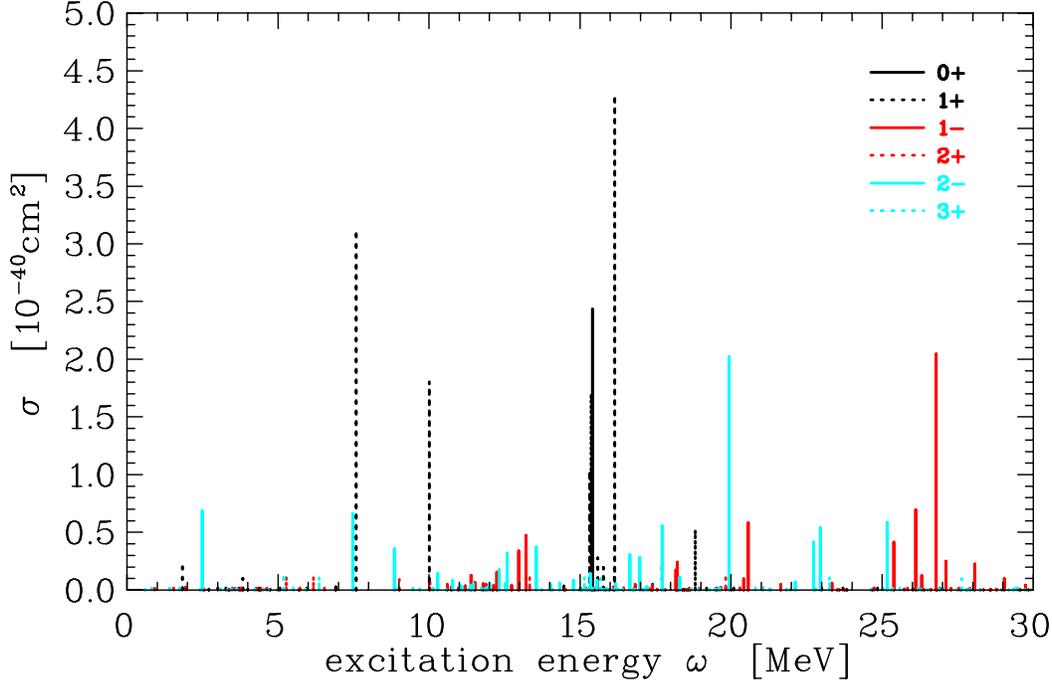}
   \caption{Multipole decomposition of the RPA response for the
            charged-current $(\nu_e,e^-)$ reaction on $^{208}$Pb 
            induced by DAR $\nu_e$ neutrinos.}
   \label{fig:pb_spectrum}
\end{figure}
The collective GT transition is found at an excitation energy 
of around $E_x=16$ MeV in $^{208}$Bi,
again close to the centroid of the experimentally observed GT strength
distribution which is at around 15.6 MeV \cite{Horen,Gaarde}. 
As has already been observed in \cite{Sagawa}, RPA calculations also
predict GT$_-$ strength at lower excitation energies, which then
correspond mainly to individual single-particle transitions. Due to
phase space, these low-lying transitions are noticeably enhanced in
neutrino-induced reactions with respect to the collective transition.
Our calculation indicates the 
low-lying GT strength to be mainly centered at around $E_x=7.5$ MeV in
$^{208}$Bi. There might be some evidence for such a transition in the
experimental (p,n) spectra on $^{208}$Pb \cite{Horen}. However, a
doubtless experimental confirmation would be quite desirable.
The first-forbidden transitions lead
mainly to $1^-$ and $2^-$ states in $^{208}$Bi. In our calculation these
transitions are fragmented over states in the energy interval between 17
MeV and 26 MeV, although we find $2^-$ strength also at rather low
excitation energies $E_x=2.5$ MeV and 7.5 MeV. 
Experimentally $2^-$ strength has been observed at $E_x=2.8$ MeV \cite{Horen}. 

To check the reliability of our approach we have performed several
additional calculations. At first
we have calculated the GT response for $^{56}$Fe within
the RPA approach. Then the GT distribution is focussed in two strong
transitions at $E_x=2$ MeV and 10.5 MeV in $^{56}$Co, corresponding to
the change of a $f_{7/2}$ neutron into  $f_{7/2}$ and $f_{5/2}$ protons,
respectively, clearly showing the inappropriate fragmentation of the GT
strength within the RPA. However, we find that this shortcoming does not
strongly influence the calculated cross section. If we correct for the
overestimation of the total RPA $S_{\beta_-}$ strength compared with the
shell model (and data), we find an RPA GT contribution to the 
$^{56}$Fe($\nu_e,e^-$)$^{56}$Co cross section in close agreement to
the shell model result (better than $3\%$). We thus conclude that our total
$^{208}$Pb($\nu_e,e^-$)$^{208}$Bi cross section, for which we could not
calculate the $\lambda^\pi=1^+$ contribution on the basis of the shell
model, is probably quite reliable.

Due to the energy and momentum-transfer
involved, muon capture is mainly sensitive to forbidden transitions
($\lambda^\pi =1^-$ and $2^-$ for $^{56}$Fe and 
$\lambda^\pi =1^+, 2^+$ and $3^+$ for $^{208}$Pb). 
We have tested our model description for forbidden transitions 
by calculating the total muon capture rates
for $^{56}$Fe and $^{208}$Pb and 
obtain results $(4.46 \cdot 10^{6}$ s$^{-1}$ 
and $16.1 \cdot 10^{6}$
s$^{-1}$) which agree rather well with experiment 
($(4.4\pm 0.1) \cdot 10^6$ s$^{-1}$ and $(13.5 \pm 0.2)  \cdot
10^6$ s$^{-1}$, respectively \cite{Suzuki}). Further details on these studies 
will be published elsewhere \cite{Kolbe00}.

The KARMEN collaboration has measured the
total $^{56}$Fe($\nu_e,e^-$)$^{56}$Co cross section 
for the DAR neutrino spectrum and obtains
$\sigma=(2.56 \pm 1.08 \pm 
0.43) \cdot 10^{-40}$ cm$^2$ \cite{Maschuw}. 
We calculate a result in close agreement
$\sigma = 2.4 \cdot 10^{-40}$
cm$^2$.
In Table 1 we have listed the partial cross sections into the 
various decay channels. 
As the isobaric analog state (IAS) at $E_x=3.5$ MeV and most of the
GT$_-$ strength resides below the particle thresholds in $^{56}$Co
(the proton and neutron thresholds are at 5.85 MeV and 10.08 MeV,
respectively), 
most of the neutrino-induced reactions on $^{56}$Fe leads to particle-bound
states, which then decay by $\gamma$ emission. 
Due to the lower threshold, neutrino-induced
 excitation of particle-unbound states in $^{56}$Co is dominantly
followed by proton decays. The rather high threshold energy (7.76 MeV)
and the larger Coulomb barrier makes decay into the $\alpha$-channel
rather unimportant.

Now we turn our discussion to $^{208}$Pb which is the shielding material 
of the LSND detector. 
The simple ($N-Z$) scaling of the Fermi and Ikeda sumrules indicates 
that the ($\nu,e^-$) cross section on $^{208}$Pb is
significantly larger than on $^{56}$Fe. The cross section is
additionally enlarged by the strong Z-dependence of the Fermi function.
In total we find that the ($\nu_e,e^-$) cross section on $^{208}$Pb is
about 15 times bigger than for $^{56}$Fe. 
Furthermore, as  the 
IAS energy and the GT$_-$ strength is
above the neutron threshold in $^{208}$Bi at 6.9 MeV, 
most of the ($\nu_e,e^-$) 
cross section leads to particle-unbound states. 
These expectations are born out by a detailed 
calculation which finds a total cross section of 
$3.62 \cdot 10^{-39}$ cm$^2$. The partial
$^{208}$Pb($\nu_e, e^- n$)$^{207}$Bi cross section dominates and
amounts to about $91 \%$ of the total cross section.
As can be seen in Table 1, the 
remaining cross section mainly goes to particle-bound levels and hence
decays by $\gamma$ emission.  

For the general reasons given above, 
our theoretical estimate for the $^{208}$Pb($\nu_e,e^-$)$^{208}$Bi 
cross section is probably quite reliable and should be useful for 
improved background simulations of the LSND detector.
It is also quite interesting to turn the problem around and ask
whether the LSND collaboration can actually measure
this cross section. To this end we have estimated the total 
number of neutrino-induced events in the lead 
shielding \cite{Atshot}
(volume $V=20$ m$^3$, density $\rho =11.3$
g/cm$^3$) of the LSND detector assuming an annual LSND neutrino flux of
$3 \cdot 10^{13}$/y. Then our $^{208}$Pb($\nu,e^-$)$^{208}$Bi 
cross section translates
into 200 000 events for the 3 year running time from 1996-98. 
In about 180 000 events a neutron is knocked out of the lead target. 
The electron will not
travel directly into the detector, but will shower in the shielding 
producing photons which in turn might reach the detector in which they
produce Compton electrons. The KARMEN collaboration has observed
this process for the  $^{56}$Fe shielding and quotes an efficiency
of their detector of $0.44 \%$. If the LSND detector has a comparable 
efficiency for this process, it should be able to observe the
$^{208}$Pb($\nu_e,e^-n$)$^{207}$Bi 
cross section where the events are most likely at
the edges. 
On the other hand, the correlated observation
of a neutron and a lepton constitutes the LSND neutrino oscillation signal.
For this reason, the LSND collaboration suppresses the events stemming from 
neutrino interactions on lead by appropriate energy and spatial cuts.
However, our calculated $^{208}$Pb($\nu_e,e^-n$) cross section might allow 
the LSND collaboration to further improve their background simulations.

The LSND oscillation experiment studies their events as 
function of energy of the outgoing lepton, 
setting cuts at 20~MeV, 36~MeV, and 53~MeV. 
We have therefore also calculated the
$^{208}$Pb($\nu,e^- n$)$^{207}$Bi cross section 
as function of the final lepton energy, which is shown in Fig.~2.
\begin{figure}[thb]
  \begin{center}
     \includegraphics[angle=90,height=9cm]{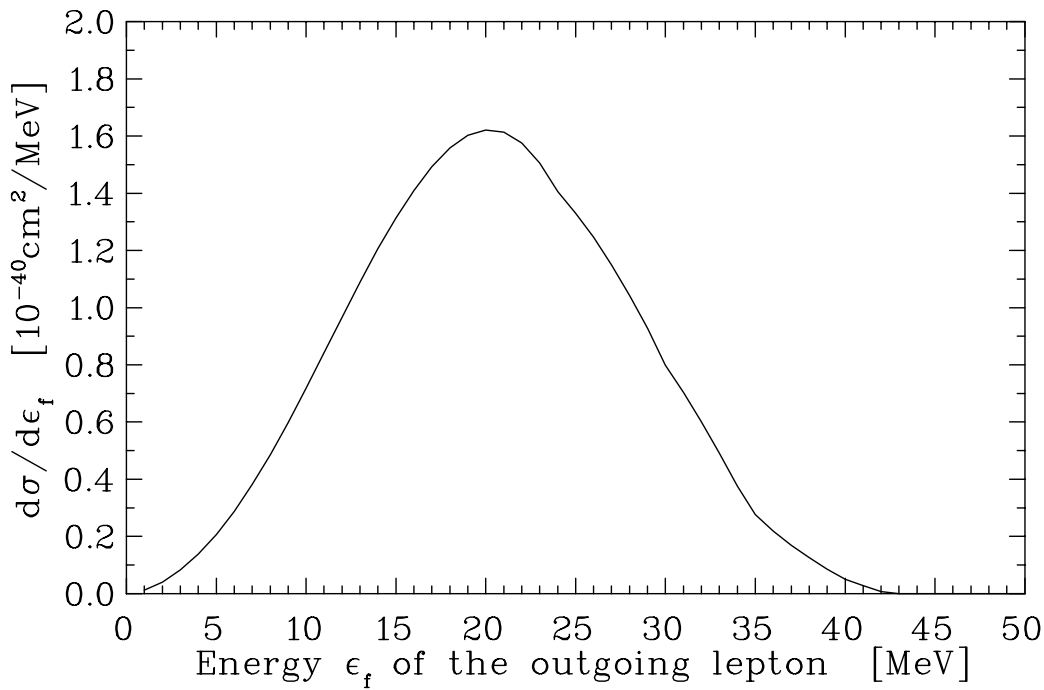}
  \end{center}
  \caption{The $^{208}$Pb($\nu_e,e^-n$)$^{207}$Bi cross section for DAR
           neutrinos as function of final electron energy.}
     \label{fig:pb_lepton}
\end{figure}
 
The LSND neutrino beam has a small admixture of $\nu_\mu$ neutrinos
stemming from pion-in-flight (DIF) decays. These neutrinos have in fact high
enough energies to significantly produce muons by the charged-current 
($\nu_\mu,\mu^-$) reaction (This beam property allowed the LSND
collaboration to measure the inclusive $^{12}$C($\nu_\mu,\mu^-$)$^{12}$N
cross section and to test universality in a neutrino
experiment on nuclei \cite{At96}). 
For the oscillation search events
stemming from the $(\nu_\mu,\mu^- n)$ reaction, 
with a possible misinterpretation of the
lepton in the final channel, are considered a possible background. For
this reason we have also calculated the total and partial ($\nu_\mu,\mu^-$)
cross sections on $^{208}$Pb for the LSND DIF $\nu_\mu$ neutrino
spectrum. The results are shown in Table 2.

We note that the `most effective' neutrino energy defined by
\begin{equation}
   {\bar E_\nu} = 
   \frac{\int E_\nu \sigma (E_\nu) dE_\nu}
   {\int  \sigma (E_\nu) dE_\nu}
\end{equation}
is larger for DIF neutrinos (${\bar E_\nu} =170$ MeV) 
than for DAR neutrinos (${\bar E_\nu} =37$ MeV). 
Thus, even if the mass difference between muon and
electron is considered, the phase space favors the reaction induced by
DIF $\nu_\mu$ neutrinos. 
Consequently the total cross section for the charged-current reaction on 
$^{208}$Pb induced by DIF $\nu_\mu$ neutrinos is larger 
(by roughly a factor 3) than induced by DAR $\nu_e$ neutrinos. 
Although the average excitation
energy in the daughter nucleus is also slightly higher for DIF $\nu_\mu$
neutrinos than for DAR $\nu_e$ neutrinos, the decay of the
particle-unbound states is still dominantly into the neutron channel.

The LSND collaboration observes candidate events which might imply 
$\nu_\mu \rightarrow \nu_e$ neutrino oscillations \cite{At98}. 
If this is the
case the DIF $\nu_\mu$ neutrinos can have changed into $\nu_e$
neutrinos before reaching the detector
now allowing for $^{208}$Pb($\nu_e,e^-$) reactions triggered
by $\nu_e$ neutrinos with a significantly higher energy. 
We have studied
the respective cross sections and have summarized them in Table 3.

For completeness, Tables 2 and 3 also list the ($\nu_\mu,\mu^-$) and
($\nu_e,e^-$) cross sections on $^{56}$Fe, in both cases calculated for a
DIF neutrino spectrum.

\subsection{Supernova neutrinos}

The observation of the neutrinos from SN1987a by the water \v{C}erenkov
detectors is generally considered as strong support that the
identification of type II supernovae as core collapse supernovae is
correct. Theoretical models predict that the proto-neutron star formed
in the center of the supernova cools by the production of neutrino
pairs, where the luminosity is approximately the same for all 3 neutrino
families. The interaction of the neutrinos with the dense surrounding,
consisting of ordinary neutron-rich matter, introduces characteristic 
differences in the neutrino distributions for the various families.
As the $\mu$ and $\tau$ neutrinos and
their antiparticles (combined referred to as $\nu_x$)
have not enough energy to
generate a muon or $\tau$ lepton, they decouple deepest in the star,
i.e. at the highest temperature, and have an average energy of ${\bar
E_\nu}=25$ MeV. As the $\nu_e$ and ${\bar \nu_e}$ neutrinos interact
with the neutron-rich matter via $\nu_e+n \rightarrow p + e^-$ and
${\bar \nu_e} + p \rightarrow n + e^+$, the ${\bar \nu_e}$ neutrinos
have a higher average energy 
(${\bar E_\nu} = 16$ MeV) than the $\nu_e$
neutrinos (${\bar E_\nu} = 11$ MeV). Clearly an observational
verification of this temperature hierarchy would establish a strong test
of our current supernova models. 

The distribution of the various supernova neutrino species is usually
described by a Fermi-Dirac spectrum
\begin{equation}
   n(E_\nu) = \frac{1}{F_2({\alpha}) T^3} 
   \frac{E_\nu^2}{\exp[(E_\nu/T)-\alpha]+1}
\end{equation}
where $T,\alpha$ are parameters fitted to numerical spectra, and
$F_2(\alpha)$ normalizes the spectrum to unit flux. The transport
calculations of Janka \cite{Janka} yield spectra with $\alpha \sim 3$
for all neutrino species. While this choice also gives good fits to the
$\nu_e$ and ${\bar \nu}_e$ spectra calculated by Wilson and Mayle
\cite{Wilson}, their $\nu_x$ spectra favor $\alpha=0$. In the following
we will present results for charged- and neutral current reactions
on $^{56}$Fe and $^{208}$Pb for both values of $\alpha$. In particular
we will include results for those 
(T,$\alpha$) values which are currently favored 
for the various neutrino types (T in MeV):
(T,$\alpha$)= (4,0) and (3,3) for $\nu_e$ neutrinos, (5,0) and (4,3) for
${\bar \nu}_e$ neutrinos and (8,0) and (6.26,3) for $\nu_x$ neutrinos.

Before discussing our neutral-current results for $^{208}$Pb we like to
present the multipole response as calculated within our RPA study. 
This is done in Fig.~3 which shows the $^{208}$Pb photoabsorption 
cross section in the upper part as well as the excitation function 
for inelastic scattering on $^{208}$Pb by neutrinos with a Fermi-Dirac 
distribution with parameters $T=8$ MeV and $\alpha=0$ in the lower part. 
The calculated photoabsorption
cross section is fragmented between 10-16 MeV excitation energy centred around
$\sim 13$ MeV. This is reasonably close to the experimental spectrum
which is centred around 13.8 MeV with a width of 3.8 MeV \cite{Brown}.
Summing over all excitation energies we obtain 3.0 MeV$\cdot$b 
for the total photoabsorption cross section, which is in  agreement
with the classical Thomas, Reiche and Kuhn sum rule value
(2.98 MeV$\cdot$b) and also lies within the range
of experimental values 
(2.9 to 4.1 MeV$\cdot$b, see table~5 of Ref.~ \cite{Ve70}).

\begin{figure}[thp]
  \begin{center}
     \includegraphics[width=14cm]{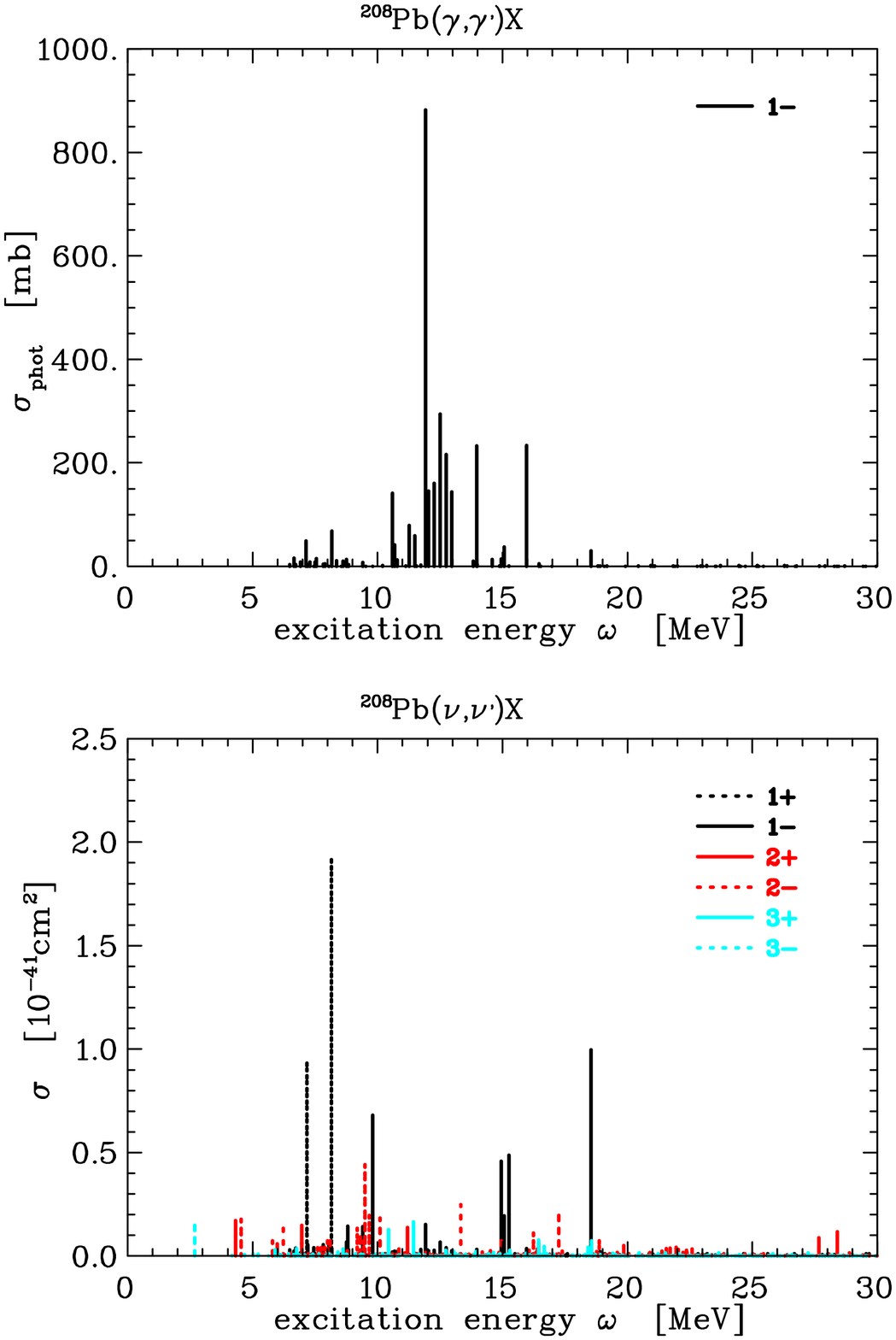}
  \end{center}
  \caption{Excitation spectrum of the $^{208}$Pb nucleus for 
           photoabsorption (upper part) in comparison to the 
           spectrum excited by neutral current neutrino scattering
           (lower part), which is decomposed into the dominant
           multipole contributions.}
     \label{fig:photNC}
\end{figure}

The lower part of 
Fig.~3 demonstrates clearly that inelastic neutrino scattering additionally
excites the spin response which is responsible for the two strong 
$J=1^-$ transitions around 10 MeV and 18 MeV. As expected from the
general effects of the residual interaction the $2^-$ part of the spin
dipole excitations is located a few MeV lower in energy than the $1^-$
strength \cite{Suzuki}. We finally note that the Gamow-Teller strength
is calculated between 7 MeV and 8 MeV, in close agreement with
the experimentally observed M1 strength.

Table 4 summarizes the total and partial cross sections for neutral
current reactions on $^{56}$Fe and $^{208}$Pb. 
For $^{56}$Fe the neutron and proton thresholds open at 11.2~MeV 
and 10.18~MeV, respectively. 
But despite the slightly higher threshold energy, the
additional Coulomb barrier in the proton channel makes the
neutron channel the dominating decay mode. 
With increasing average neutrino energies the total cross section grows. 
But this increase is
noticeably weaker than for the nuclei $^{12}$C and $^{16}$O. This is
related to the isovector dominance of the neutrino-induced reactions.
In the $T=0$ nuclei $^{12}$C and $^{16}$O  
inelastic neutrino scattering has to overcome a rather large threshold
to reach the $T=1$ excitation spectrum in the nuclei making the cross
section rather sensitive to the neutrino spectrum.

The total and partial cross sections for charged current 
($\nu_e,e^-$) and (${\bar \nu}_e,e^+$) reactions on
$^{56}$Fe and $^{208}$Pb are listed in Table 5.
As the average energy for supernova $\nu_e$ neutrinos
(${\bar E_\nu}  \approx 11$ MeV)
is less than for  DAR neutrinos
(${\bar  E_\nu}  \approx 37$ MeV), the total cross sections are
significantly smaller for supernova (i.e. (T,$\alpha$)=(4,0) or (3,3))
neutrinos. Relatedly the low-energy excitation spectrum is stronger weighted
by phase space. Hence, the $\nu_e$-induced reaction on $^{56}$Fe leads
dominantly to particle-bound states ($\sim 60\%$) and therefore decays
by $\gamma$ emission. 
As for DAR neutrinos, the strongest decay mode for $\nu_e$-induced
reactions on $^{208}$Pb is given by the neutron channel.

As lead is discussed as material for potential supernova neutrino
detectors (like LAND and OMNIS), the relevant neutrino-induced reactions
on $^{208}$Pb have been estimated previously. The first work, performed
in \cite{Land}, has been criticized and improved in \cite{Fuller}.
These authors estimated the allowed transitions to the charged-current
and neutral-current cross sections empirically using data from (p,n)
scattering and from the M1 response to fix the Gamow-Teller
contributions to the cross section. We note that these data place the
GT$_-$ strength in one resonance centered just above the 2n
threshold. Low-lying  GT$_-$ transitions,
as indicated by the present RPA calculation, have not been considered
in \cite{Fuller}. 
Ref. \cite{Fuller} completed their cross section estimates by
 calculating the first-forbidden contributions on the basis
of the Goldhaber-Teller model.

Although the total charged-current $^{208}$Pb($\nu_e,e^-$)$^{208}$Bi cross
section is strongly constrained by sumrules and our calculation
as well as the work of Ref.~\cite{Fuller}  reproduce 
the energies of the IAS state and the main GT resonance,
our results clearly deviate with increasing neutrino energies from
the calculation of Ref.~\cite{Fuller}. 
For $\nu_e$ neutrinos with a (T,$\alpha$)=(3,3) Fermi-Dirac distribution 
our  cross section ($1.6 \times 10^{-40}$ cm$^2$)
is in rough agreement with the one obtained in \cite{Fuller}. 
(As \cite{Fuller} does not give
the cross section for a (T=3,$\alpha$=3) spectrum, we have estimated it
from the cross sections given at neighboring temperatures taken from
Table I of \cite{Fuller}.)
But with increasing neutrino energies 
our calculated cross sections become significantly smaller
than the estimate given in \cite{Fuller}, and for a $\nu_e$ spectrum with
(T,$\alpha$)=(8,0) our value ($25 \times 10^{-40}$ cm$^2$)
is about $55\%$ smaller than the estimate by Ref.~\cite{Fuller}
($58 \times 10^{-40}$ cm$^2$). 
For the latter neutrino spectrum the cross section is dominated
by forbidden transitions, and the observed difference might reflect the
uncertainties of the Goldhaber-Teller model to describe this
response. 

For the total neutral-current cross sections on $^{208}$Pb
the estimates in \cite{Fuller} are noticeably larger 
than our results (by factors in the range 2--3 for the various
Fermi-Dirac spectra) for all energies.
As pointed out by Haxton \cite{Wo90} 
the total ($\nu,\nu^\prime$) cross sections on nuclei induced by
supernova neutrinos
with high energetic Fermi-Dirac distributions 
follow a simple rule of thumb: 
\begin{equation}
   \sigma (\nu,\nu^\prime) = c(T,\alpha) \cdot A \cdot 10^{-42} {\rm cm}^2  .
   \label{RUTH}
\end{equation}   
The proportionality factor  depends  on the 
parameters of the Fermi-Dirac spectrum. From RPA studies one finds 
$c(T,\alpha) \approx 0.7-0.9$ for $T=8$ MeV and $\alpha=0$ \cite{Qian,Hektor},
while the proportionality factor is slightly smaller for closed-shell
nuclei. 
We note that our present results fit well into the expected systematics:
$c(T=8 MeV, \alpha=0) = 0.77$ for $^{56}$Fe (open shell) 
and 0.67 for $^{208}$Pb (closed shell).

Besides detecting a supernova neutrino signal, modern detectors should
also have a `neutrino spectroscopy ability', i.e. it is desirable to
assign observed events to the neutrino type which has triggered it.
Detectors like LAND and OMNIS will observe the neutrons produced by
neutrino-induced reactions on $^{208}$Pb. 
An obvious neutrino signal then is the total count rate. However, as
already pointed out in \cite{Fuller}, the total neutron count rate in a
lead detector does not allow to distinguish between events triggered by
$\nu_e$ neutrinos and $\nu_x$ neutrinos. We confirm this argument as our
total ($\nu_e,e^- n$) cross section (e.g. for (T,$\alpha$)=(4,0) it is
$2.3 \times 10^{-40}$ cm$^2$) is quite similar to the neutral current
cross section (for (T,$\alpha$)=(8,0) neutrinos we find $1.4 \times 10^{-40}$
cm$^2$ per neutrino family). The situation is, however, different
for $^{56}$Fe. Here we find, for the same neutrino spectra as above,
that the  total neutron counting rate in the neutral-current 
reaction is about 30~times larger than for the
charged-current reaction. If we consider that supernova $\nu_x$
neutrinos comprise 4 neutrino types with about the same spectrum, the
neutron response of a $^{56}$Fe detector to supernova neutrinos is expected to
be dominated by neutral current events caused by $\nu_x$ neutrinos.

The differences in the ratios for neutral- and charged current
neutron yields again reflect the more general tendency that neutral-current
cross sections for supernova $\nu_x$ neutrinos scale approximately with
the mass number $A$ of the target, while the charged-current cross
sections for supernova $\nu_e$ neutrinos depends on the $N-Z$ neutron
excess of the target via the Fermi and Ikeda sumrules (e.g.
\cite{Hektor}). 
This suggests \cite{Fuller} that neutrino
detectors which can only determine total neutron counting rates 
can have supernova neutrino spectroscopy ability if they are made of various
materials with quite different $Z$ values as the ratio of neutral- to
charged-current cross sections is quite sensitive to the charge number of
the detector material. Of course, it is then necessary to assign
observed events to the detector material.

Neutrino detectors of large size will probably not be build from
isotopically enriched iron or lead, because the costs will be very high.
Therefore, in principle, in addition to $^{56}$Fe (91.75\% natural abundance)
and $^{208}$Pb (52.4\%), also cross sections for neutrino induced 
reactions on the other stable isotopes $^{54}$Fe (5.85\%), $^{57}$Fe (2.12\%),
$^{58}$Fe (0.28\%), $^{206}$Pb (24.1\%), $^{207}$Pb (22.1\%), 
$^{204}$Pb (1.4\%) are needed.
But from the rule of thumb (Eq.~\ref{RUTH}) we can already conclude that
the isotope effect on the neutral-current cross sections will be small.
This has been confirmed for the iron isotope chain $^{52-58}$Fe 
 within a recent shell model plus RPA approach which finds 
less than 16\% deviation from the simple scaling rule 
(Eq. \ref{RUTH}) \cite{To00}.
Contrary the isotope effect on the charged-current cross sections will be 
strong, because they dominantly scale with (N-Z) like mentioned above 
via the Fermi and Ikeda sumrules. This is again confirmed in the 
shell model plus RPA study which finds less than $10\%$ deviation in the
charged-current cross sections for $T=4$ MeV and $\alpha=0$ neutrinos
from the simple $(N-Z)$ scaling \cite{To00}.
We expect that the rule of thumb (Eq. (4)) and the $(N-Z)$ scaling is
also valid for the neutral-current and charged-current reactions on
$^{208}$Pb, respectively.
This provides then a simple scheme to estimate the charged-current
cross sections for the other lead and iron isotopes.

Both the LAND and the OMNIS detectors will also be capable of detecting
the neutron energy spectrum following the decay of states in the
daughter nucleus after excitation by charged- and neutral-current
neutrino reactions. We have calculated the relevant neutron energy
spectra for both possible detector materials, $^{56}$Fe and $^{208}$Pb.
To this end we have used the statistical model code SMOKER iteratively
by following the decay of the daughter states after the first particle
decay. We have kept book of the neutron energies produced in these
(sequential) decays and have binned them in 500-keV bins. The neutron
energy spectra obtained this way are shown in Figs. 4--7. The
calculations have been performed for different neutrino spectra which
also allows one to study the potential sensitivity of the detectors if
neutrino oscillations occur.

For the charged- and neutral-current reactions on $^{56}$Fe the response
is mainly below the 2n-threshold. Most of the Gamow-Teller distribution
is below the neutron-threshold, as is the IAS in the charged-current
reaction. 
The neutron energy spectrum of the $^{56}$Fe($\nu_e,e^- n$) reaction is
shown in Fig. 4.
\begin{figure}[thb]
  \begin{center}
     \includegraphics[angle=90,height=9cm]{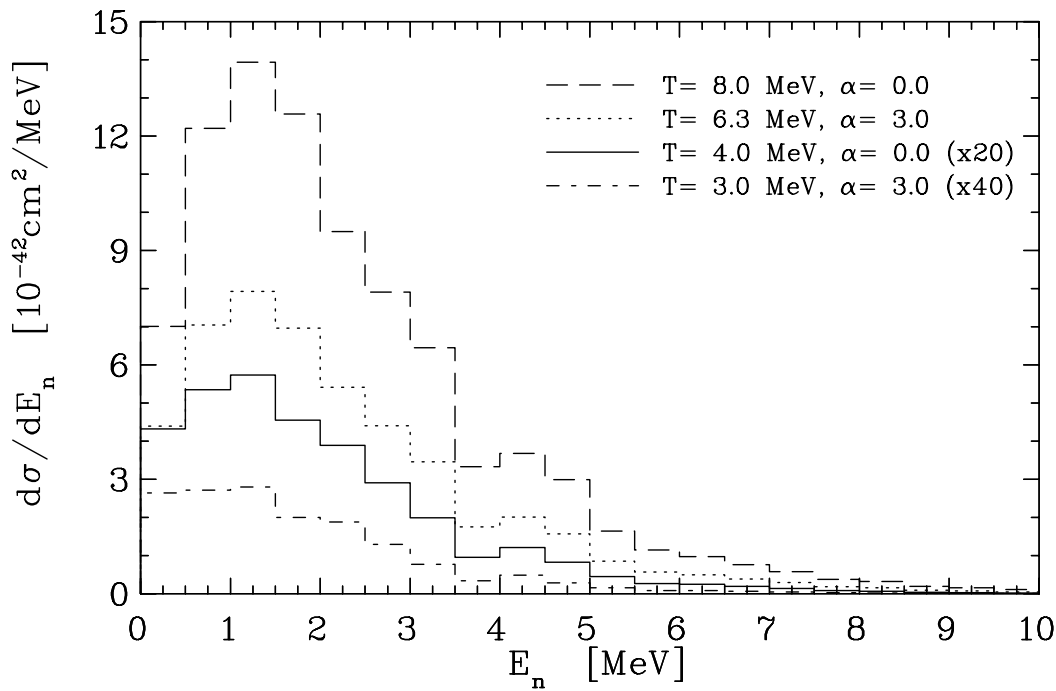}
  \end{center}
  \caption{Neutron energy spectrum produced by the charged-current
           ($\nu_e,e^-$) reaction on $^{56}$Fe. 
           The calculation has been performed for different 
           supernova neutrino spectra characterized by 
           the parameters (T,$\alpha$).
           Note that the cross sections for (T,$\alpha$)=(4,0) 
           and (3,3) neutrinos have been scaled by factors 20 and 40, 
           respectively.}
\end{figure} 
The spectrum is rather structureless with a broad peak centred around neutron
energies $E_n=1 - 1.5$ MeV and basically reflects the
GT$_-$ distribution above the neutron threshold of 10.08 MeV.
The respective neutron spectrum for the neutral current reaction is
shown in Fig. 5.
\begin{figure}[thb]
  \begin{center}
     \includegraphics[angle=90,height=9cm]{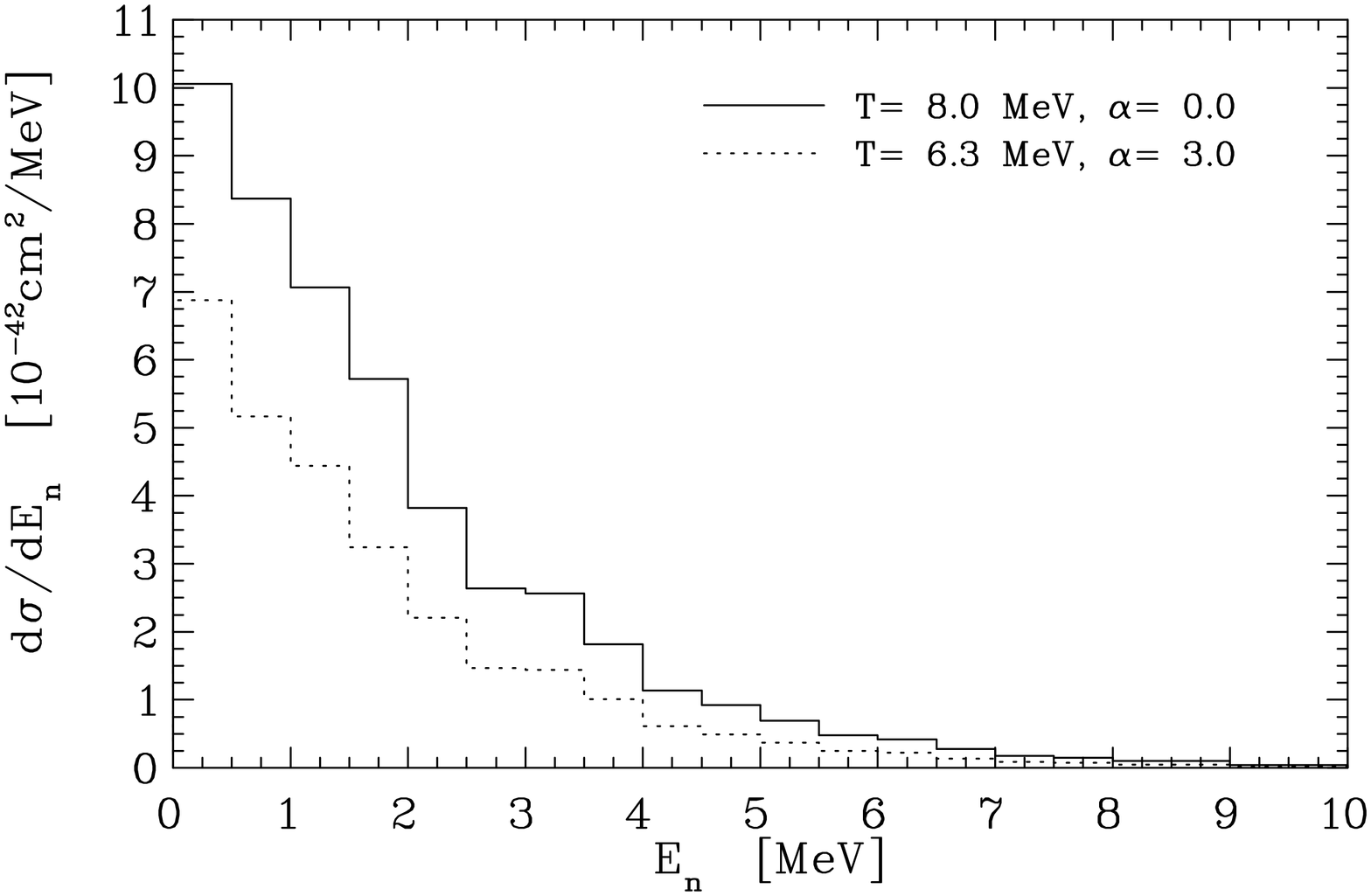}
  \end{center}
  \caption{Neutron energy spectrum produced by the neutral-current
           ($\nu,\nu'$) reaction on $^{56}$Fe. 
           The calculation has been performed for different supernova 
           neutrino spectra characterized by the parameters (T,$\alpha$).}
\end{figure} 
The spectrum is composed by several (mainly first-forbidden) transitions
which combined lead to a rather smooth neutron energy distribution.
We note that the GT distribution is taken from the shell model
calculation and leads to a rather broad neutron spectrum.

The neutron spectrum for the charged current reaction on $^{208}$Pb is
dominated by the Fermi transition to the IAS and by the GT$_-$
transitions. To understand the neutron spectrum we have to consider the
neutron threshold energies for one-neutron decay (6.9 MeV) and for
two-neutron decay (14.98 MeV) in $^{208}$Bi. Hence the IAS and 
the collective
GT resonance (with an excitation energy of about 16 MeV) will decay
dominantly by 2n emission, while the low-lying GT$_-$ resonance 
at $E_x=7.6$ MeV decays by
the emission of one neutron. This has significant consequences for the
neutron spectrum. In the 2-neutron decay the available energy is shared
between the two emitted particles, leading 
to a rather broad and
structureless neutron energy distribution. 
As can be seen in Fig. 6, this broad structure is
overlaid with a peak at neutron energy around $E_n=1$ MeV caused by the
one-neutron decay of the lower GT$_-$ transition.
\begin{figure}[thb]
  \begin{center}
     \includegraphics[angle=90,height=9cm]{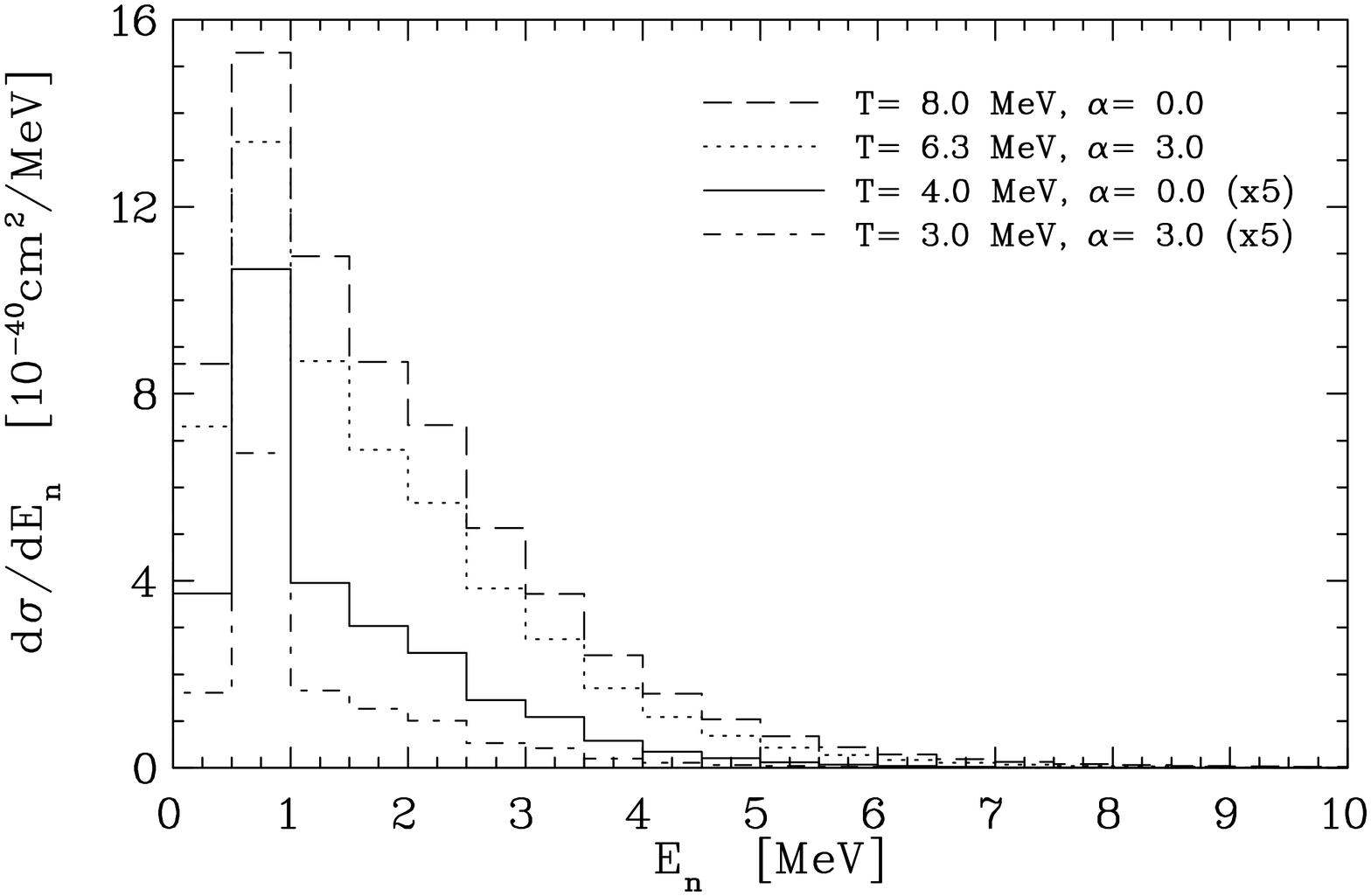}
  \end{center}
  \caption{Neutron energy spectrum produced by the charged-current
           ($\nu_e,e^-$) reaction on $^{208}$Pb. 
           The calculation has been performed for different supernova 
           neutrino spectra characterized by the parameters (T,$\alpha$).
           Note that the cross sections for (T,$\alpha$)=(4,0) 
           and (3,3) neutrinos have been scaled by a factor 5.}
\end{figure}
We expect that due to fragmentation, not properly described in our RPA
calculation, the width of this peak might be broader than the 
0.5~MeV-binning which we have assumed in Fig.~6.
We note that the relative height of the peak compared with the broad
structure stemming from the 2n-emission is more pronounced for the
(T,$\alpha$)=(4,0) neutrino distribution than for a potential
(T,$\alpha$)=(8,0) $\nu_e$ spectrum as it might arise after complete
$\nu_e \leftrightarrow \nu_\mu$ oscillations.

Fig.~7 shows the neutron energy spectrum for the neutral-current
reactions on $^{208}$Pb. 
\begin{figure}[thb]
  \begin{center}
     \includegraphics[angle=90,height=9cm]{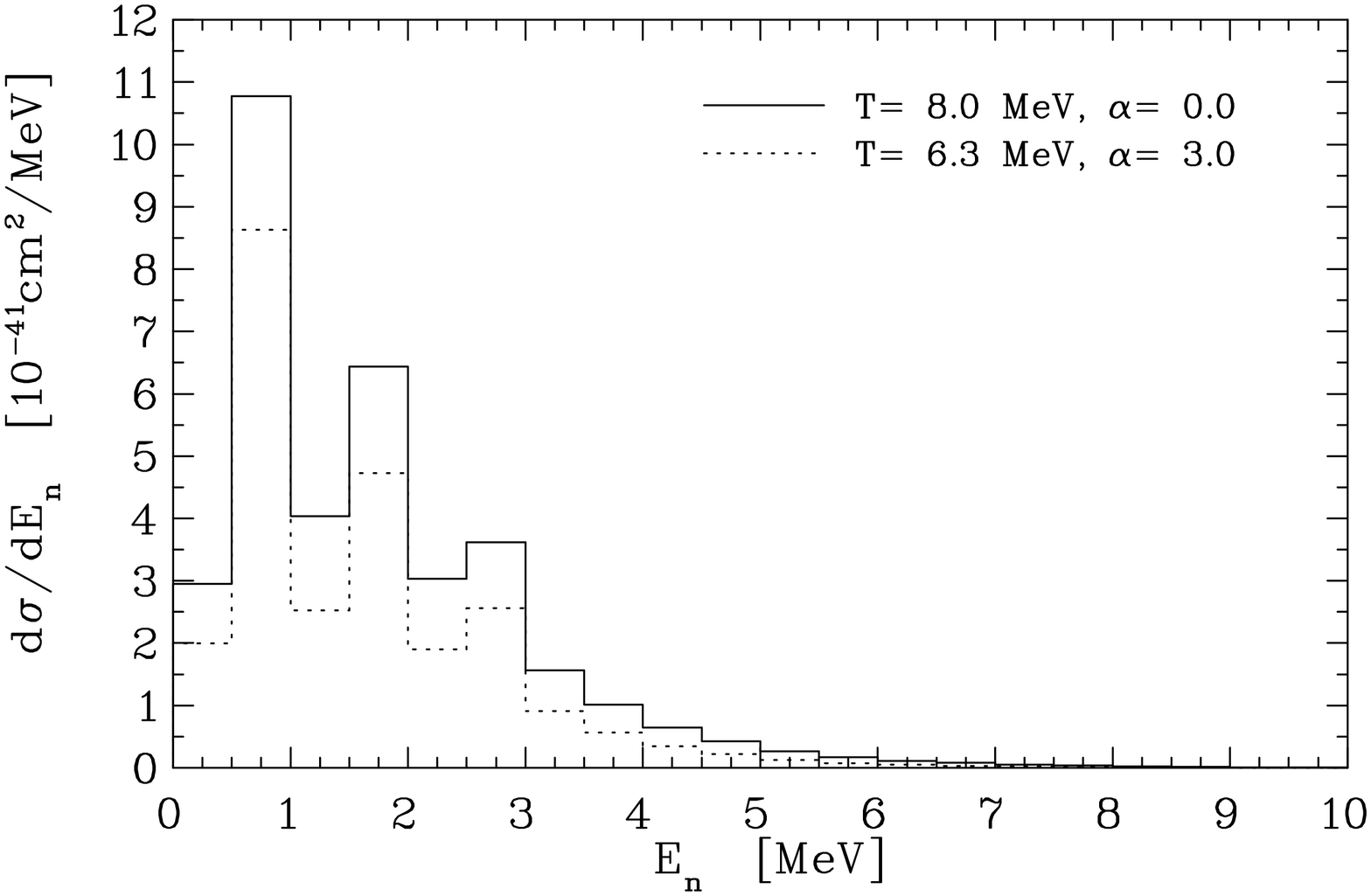}
  \end{center}
  \caption{Neutron energy spectrum produced by the neutral-current
           ($\nu,\nu'$) reaction on $^{208}$Pb. 
           The calculation has been performed for different supernova 
           neutrino spectra characterized by the parameters (T,$\alpha$).}
\end{figure}
Our RPA response places the strong GT
transitions around the neutron threshold (at 7.37 MeV), while the
first-forbidden transitions are split into several transitions between
the excitation energies 9 MeV and 18 MeV. In particular, the two strong
$1^-$ resonances at around 15 MeV and 18 MeV are above the 2-neutron
threshold at 14.12~MeV and their decay leads, for the same reasons as 
given above for
the charged-current reaction, two a rather broad neutron energy
spectrum. Several transitions above the one-neutron threshold superimpose in
our RPA neutron spectrum this broad structure and lead to 
rather pronounced peaks. But nucleon-nucleon correlations beyond the RPA
will induce a stronger fragmentation which will smear out these peaks.
We expect therefore that the neutral-current neutron energy spectrum
will be rather broad and structureless.

An exciting question is whether supernova neutrino detectors have the
ability to detect neutrino oscillations. This can be achieved by a 
suited signal which allows
to distinguish between charged-current and neutral-current events and
which is quite sensitive to the neutrino distribution. 
It is hoped for that the detectors
OMNIS and LAND have such an ability. However, as has been shown in
\cite{Fuller}, the total neutron counting rate is by itself not a suited
mean to detect neutrino oscillations, even if results from various
detectors with different material (hence different ratios of 
charged-to-neutral current cross sections, as discussed above) are combined.
In Ref. \cite{Fuller} it is pointed out that in the case of $^{208}$Pb an
attractive signal might emerge. 
Due to the fact that the IAS and large
portions of the GT$_-$ strength resides in $^{208}$Bi just above the
2-neutron emission threshold, Fuller {\it et al.} discuss that the
2-neutron emission rate is both, flavor-specific and very sensitive to
the  temperature of the $\nu_e$ distribution.
To quantify this argument we have calculated the cross sections for the
$^{208}$Pb($\nu_e, e^- 2n$)$^{206}$Bi reaction in our combined model of
RPA for the neutrino-induced response and statistical model for the
decay of the daughter states. We find the partial cross sections of
$43.9 \times 10^{-42}$ cm$^2$ and
$13.0 \times 10^{-42}$ cm$^2$ for $\nu_e$ neutrinos with 
(T,$\alpha$)=(4,0) and (3,3) Fermi-Dirac distributions. As pointed out
in \cite{Fuller} these cross sections increase significantly if neutrino
oscillations occur. For example, we find for total $\nu_e
\leftrightarrow \nu_\mu$ oscillations partial 2n cross sections of
$1053 \times 10^{-42}$ cm$^2$ and
$742 \times 10^{-42}$ cm$^2$ (for neutrino distributions with
parameters (T,$\alpha$)= (8,0) and (6.26,3), respectively). We remark
that these numbers will probably be reduced, if correlations beyond the
RPA are taken into account, as part of the GT$_-$ distribution might be
shifted below the 2n-threshold.

As pointed out above, also portions of the neutral-current excitation
spectrum are above the respective 2n-emission threshold. This decay will
compete with the one stemming from the charged-current reaction and
hence will reduce the flavor-sensitivity of the signal. We have
therefore also calculated the $^{208}$Pb($\nu,\nu' 2n$)$^{206}$Pb cross
sections and find 
$41.3 \times 10^{-42}$ cm$^2$ and
$23.5 \times 10^{-42}$ cm$^2$ (for neutrino distributions with
(T,$\alpha$)= (8,0) and (6.26,3), respectively and averaged over
neutrinos and antineutrinos). Thus, if no neutrino oscillations occur
the combined 2n-signal resulting from neutral-current reactions 
for the 4 $\nu_x$
neutrino types is larger than the one from the charged-current
reactions. However, if neutrino oscillations occur the neutral-current
signal is unaffected while the charged-current signal is drastically
enhanced. Thus, our calculations support the suggestions of Ref.
\cite{Fuller} that the 2n-signal for $^{208}$Pb detectors might be an
interesting neutrino oscillation signal. However, 
our calculations also indicate
that, for an analysis of the potential observation of the signal,
2-neutron emission from neutral-current events have to be accounted for
as well.

Finally, as the predicted energy spectra of neutrinos 
from supernovae change with time  and furthermore can be 
affected in a variety of ways
(especially oscillation scenarios), Table~6 lists the cross sections for 
($\nu_e,e^-$)- and ($\nu,\nu'$)-scattering 
on $^{56}$Fe and $^{208}$Pb as a function of neutrino energy.

\section{Conclusions}

We have studied the charged- and neutral current reactions on $^{56}$Fe
and $^{208}$Pb which are the shielding materials for current
accelerator-based neutrino experiments like LSND and KARMEN and the
material for proposed supernova neutrino detectors like LAND
and OMNIS.

Our calculations for $^{56}$Fe are performed within a model which uses
the interacting shell model to determine the Gamow-Teller response and
the RPA for forbidden transitions. For $^{208}$Pb the complete nuclear
response is evaluated within the RPA model. The correct
momentum-dependence of the various multipole-operators is considered.
This leads to a reduction of the cross sections, compared to
calculations performed at $q=0$
due to destructive interference with `higher-order' multipole operators.

At first we have calculated the total cross sections and the partial
cross sections for spallating a neutron from the target for
muon-decay-at-rest neutrinos. Additionally we have evaluated the 
charged-current cross
section on $^{208}$Pb as a function of final lepton energy. All these
quantities are expected to allow for (even) more reliable background 
simulations for the LSND and KARMEN detectors. As the LSND 
collaboration might have
observed a neutrino-oscillation signal we have also calculated the
various cross sections on $^{56}$Fe and $^{208}$Pb for pion-in-flight-decay
neutrinos as they comprise a small admixture of $\nu_\mu$ neutrinos in
the LSND beam.

Detecting supernova neutrinos is generally considered an important test
of theoretical models for core-collapse supernovae. OMNIS and LAND are
two proposed detectors, consisting of lead and possibly iron, which will
have the capability to count the total rate of neutrons produced by
neutrino reactions in the detector and further to detect the related neutron
energy spectrum.  For $^{56}$Fe the decay is mainly by emission of one
neutron. Nevertheless the neutron energy spectrum is rather broad and
structureless following both charged- and neutral-current excitations.

For $^{208}$Pb the situation is different as a significant portion of
the charged-current response (and also of the neutral-current response)
is above the 2n-threshold. As the two neutrons share the available
decay energy this leads to a rather broad neutron spectrum. For the
charged-current reaction we predict that this broad pattern is
superimposed by a peak structure, due to a yet
unobserved Gamow-Teller transition at lower energies.
We find that the height of this peak relative to the broad structure is
more pronounced for `ordinary' $\nu_e$ supernova neutrinos than for a
$\nu_e$ neutrino spectrum arising after $\nu_\mu \rightarrow \nu_e$
oscillations. Another possible oscillation signal for a $^{208}$Pb
detector is the emission rate of 2 neutrons, as suggested by Fuller,
Haxton and McLaughlin. We have quantitatively confirmed the argument of
these authors and have also calculated the 2-neutron emission rate for
the neutral-current reaction which has to be considered if, in the event
of a nearby supernova, the 2-neutron emission signal would be observed
and analyzed for oscillation information.

\acknowledgments{We thank G. Drexlin and M. Steidl from the KARMEN-Group 
for stimulating and helpful discussions. We are also grateful to G.
Mart\'{\i}nez-Pinedo for his help with the shell model calculations.
The work has partly been supported by a grant of the Danish Research Council.}

\renewcommand\topfraction{.15}
\renewcommand\textfraction{.85}

\begin{table}[thb]
  \begin{center}
    \begin{tabular}{|l|c|} \hline \hline 
      \rule[-1ex]{0em}{4ex} neutrino reaction & cross section \\ 
       \hline \hline
       \nucleus{\rm Fe}{56}($\nu_e,e^-\gamma$)\nucleus{\rm Co}{56} &
         1.25 ( 2) \\ 
       \nucleus{\rm Fe}{56}($\nu_e,e^-$n)\nucleus{\rm Co}{55} &
         3.33 ( 1) \\ 
       \nucleus{\rm Fe}{56}($\nu_e,e^-$p)\nucleus{\rm Fe}{55} &
         7.83 ( 1) \\ 
       \nucleus{\rm Fe}{56}($\nu_e,e^-\alpha$)\nucleus{\rm Mn}{52} &
         3.52 ( 0) \\ 
       \nucleus{\rm Fe}{56}($\nu_e,e^-$)X &
         2.40 ( 2) \\ 
       \hline
       \nucleus{\rm Pb}{208}($\nu_e,e^-\gamma$)\nucleus{\rm Bi}{208} &
         3.24 ( 2) \\
       \nucleus{\rm Pb}{208}($\nu_e,e^-$n)\nucleus{\rm Bi}{207} &
         3.29 ( 3) \\
       \nucleus{\rm Pb}{208}($\nu_e,e^-$p)\nucleus{\rm Pb}{207} &
         4.77 (-1) \\
       \nucleus{\rm Pb}{208}($\nu_e,e^-\alpha$)\nucleus{\rm Tl}{204} &
         1.01 ( 0) \\
       \nucleus{\rm Pb}{208}($\nu_e,e^-$)X &
         3.62 ( 3) \\
       \hline\hline
    \end{tabular}
  \end{center}
  \caption{Total cross sections for charged current 
           neutrino scattering on nuclei for electron neutrinos from
           pion-decay-at-rest. The 
           cross sections are given in units of $10^{-42} {\rm cm}^2$,
           exponents are given in parenthesis.}
\end{table}

\begin{table}[thb]
  \begin{center}
    \begin{tabular}{|l|c|} \hline \hline 
      \rule[-1ex]{0em}{4ex} neutrino reaction & cross section \\ 
       \hline \hline
       \nucleus{\rm Fe}{56}($\nu_\mu,\mu^-\gamma$)\nucleus{\rm Co}{56} &
         2.24 ( 2) \\ 
       \nucleus{\rm Fe}{56}($\nu_\mu,\mu^-$n)\nucleus{\rm Co}{55} &
         6.62 ( 2) \\ 
       \nucleus{\rm Fe}{56}($\nu_\mu,\mu^-$p)\nucleus{\rm Fe}{55} &
         1.33 ( 3) \\ 
       \nucleus{\rm Fe}{56}($\nu_\mu,\mu^-\alpha$)\nucleus{\rm Mn}{52} &
         2.23 ( 2) \\ 
       \nucleus{\rm Fe}{56}($\nu_\mu,\mu^-$)X &
         2.44 ( 3) \\ 
       \hline
       \nucleus{\rm Pb}{208}($\nu_\mu,\mu^-\gamma$)\nucleus{\rm Bi}{208} &
         1.23 ( 3) \\
       \nucleus{\rm Pb}{208}($\nu_\mu,\mu^-$n)\nucleus{\rm Bi}{207} &
         1.02 ( 4) \\
       \nucleus{\rm Pb}{208}($\nu_\mu,\mu^-$p)\nucleus{\rm Pb}{207} &
         2.89 ( 0) \\
       \nucleus{\rm Pb}{208}($\nu_\mu,\mu^-\alpha$)\nucleus{\rm Tl}{204} &
         3.31 ( 1) \\
       \nucleus{\rm Pb}{208}($\nu_\mu,\mu^-$)X &
         1.15 ( 4) \\
       \hline\hline
    \end{tabular}
  \end{center}
  \caption{Total cross sections for charged current 
           neutrino scattering on nuclei for muon neutrinos with the
           LSND pion-decay-in-flight spectrum. 
           The cross sections are given in units of $10^{-42} {\rm cm}^2$,
           exponents are given in parenthesis.}
\end{table}

\begin{table}[thb]
  \begin{center}
    \begin{tabular}{|l|c|} \hline \hline 
      \rule[-1ex]{0em}{4ex} neutrino reaction & cross section \\ 
       \hline \hline
       \nucleus{\rm Fe}{56}($\nu_e,e^-\gamma$)\nucleus{\rm Co}{56} &
         5.80 ( 2) \\ 
       \nucleus{\rm Fe}{56}($\nu_e,e^-$n)\nucleus{\rm Co}{55} &
         1.91 ( 3) \\ 
       \nucleus{\rm Fe}{56}($\nu_e,e^-$p)\nucleus{\rm Fe}{55} &
         3.84 ( 3) \\ 
       \nucleus{\rm Fe}{56}($\nu_e,e^-\alpha$)\nucleus{\rm Mn}{52} &
         6.48 ( 2) \\ 
       \nucleus{\rm Fe}{56}($\nu_e,e^-$)X &
         6.98 ( 3) \\ 
       \hline
       \nucleus{\rm Pb}{208}($\nu_e,e^-\gamma$)\nucleus{\rm Bi}{208} &
         2.75 ( 3) \\
       \nucleus{\rm Pb}{208}($\nu_e,e^-$n)\nucleus{\rm Bi}{207} &
         3.49 ( 4) \\
       \nucleus{\rm Pb}{208}($\nu_e,e^-$p)\nucleus{\rm Pb}{207} &
         1.00 ( 1) \\
       \nucleus{\rm Pb}{208}($\nu_e,e^-\alpha$)\nucleus{\rm Tl}{204} &
         1.12 ( 2) \\
       \nucleus{\rm Pb}{208}($\nu_e,e^-$)X &
         3.78 ( 4) \\
       \hline\hline
    \end{tabular}
  \end{center}
  \caption{Total cross sections for charged current $(\nu_e,e^-)$ 
           neutrino scattering on $^{56}$Fe and $^{208}$Pb nuclei for 
           electron neutrinos with the LSND 
           pion-decay-in-flight neutrino spectrum. 
           The cross sections are given in units of $10^{-42} {\rm cm}^2$,
           exponents are given in parenthesis.}
\end{table}

\begin{table}[thb]
   \begin{center}
   \begin{tabular}{|l|c|c|c|c|c|c|c|} \hline \hline 
      \rule[-1ex]{0em}{4ex} { \hspace{1.8em}} ($T, \alpha$) & 
       (4,0) & (6,0) & (8,0) & (10,0) & (3,3) & (4,3) & (6.26,3) \\ 
       \hline \hline 
       \nucleus{\rm Fe}{56}($\nu,\nu^{\prime} \gamma$)\nucleus{\rm Fe}{56} &
        2.9 ( 0)& 9.3 ( 0)& 1.9 ( 1)& 3.0 ( 1)& 1.9 ( 0)& 5.0 ( 0)& 1.7 ( 1)\\
       \nucleus{\rm Fe}{56}($\nu,\nu^{\prime}$n)\nucleus{\rm Fe}{55} &
        7.1 (-1)& 5.9 ( 0)& 2.1 ( 1)& 4.9 ( 1)& 2.3 (-1)& 1.3 ( 0)& 1.3 ( 1)\\
       \nucleus{\rm Fe}{56}($\nu,\nu^{\prime}$p)\nucleus{\rm Mn}{55} &
        5.6 (-2)& 6.8 (-1)& 3.1 ( 0)& 8.7 ( 0)& 1.3 (-2)& 1.1 (-1)& 1.6 ( 0)\\
       \nucleus{\rm Fe}{56}($\nu,\nu^{\prime} \alpha$)\nucleus{\rm Cr}{52} &
        9.4 (-3)& 1.2 (-1)& 5.5 (-1)& 1.6 ( 0)& 2.1 (-3)& 1.8 (-2)& 2.8 (-1)\\
       \nucleus{\rm Fe}{56}($\nu,\nu^{\prime}$)X &
        3.7 ( 0)& 1.6 ( 1)& 4.3 ( 1)& 9.0 ( 1)& 2.1 ( 0)& 6.4 ( 0)& 3.2 ( 1)\\
       \hline
       \nucleus{\rm Pb}{208}($\nu,\nu^{\prime} \gamma$)\nucleus{\rm Pb}{208} &
         3.6 ( 0)& 1.2 ( 1)& 2.7 ( 1)& 4.8 ( 1)& 2.4 ( 0)& 6.1 ( 0)& 2.2 ( 1)\\
       \nucleus{\rm Pb}{208}($\nu,\nu^{\prime}$n)\nucleus{\rm Pb}{207} &
         1.1 ( 1)& 5.0 ( 1)& 1.4 ( 2)& 2.8 ( 2)& 5.8 ( 0)& 1.9 ( 1)& 1.0 ( 2)\\
       \nucleus{\rm Pb}{208}($\nu,\nu^{\prime}$p)\nucleus{\rm Tl}{207} &
         2.3 (-5)& 5.3 (-4)& 3.8 (-3)& 1.5 (-2)& 4.0 (-6)& 4.4 (-5)& 1.4 (-3)\\
       \nucleus{\rm Pb}{208}($\nu,\nu^{\prime} \alpha$)\nucleus{\rm Hg}{204} &
         1.2 (-4)& 4.7 (-3)& 4.7 (-2)& 2.3 (-1)& 1.2 (-5)& 2.2 (-4)& 1.3 (-2)\\
       \nucleus{\rm Pb}{208}($\nu,\nu^{\prime}$)X &
         1.4 ( 1)& 6.2 ( 1)& 1.6 ( 2)& 3.3 ( 2)& 8.1 ( 0)& 2.5 ( 1)& 1.2 ( 2)\\
       \hline\hline
    \end{tabular}
  \end{center}
  \caption{Total cross sections for neutral current 
           neutrino scattering on nuclei for different neutrino energy
           spectra represented as Fermi-Dirac distributions. 
           The cross sections are given in units of $10^{-42} {\rm cm}^2$ 
           and are averaged over neutrinos and antineutrinos.}         
\end{table}

\begin{table}[thb]
  \begin{center}
    \begin{tabular}{|l|c|c|c|c|c|c|c|} \hline \hline 
      \rule[-1ex]{0em}{4ex} { \hspace{1.8em}} ($T, \alpha$) &  
       (4,0) & (6,0) & (8,0) & (10,0) & (3,3) & (4,3) & (6.26,3) \\
       \hline \hline
       \nucleus{\rm Fe}{56}($\nu_e,e^-\gamma$)\nucleus{\rm Co}{56} &
         9.8 ( 0)& 3.2 ( 1)& 6.4 ( 1)& 1.0 ( 2)& 6.5 ( 0)& 1.7 ( 1)& 5.9 ( 1)\\
       \nucleus{\rm Fe}{56}($\nu_e,e^-$n)\nucleus{\rm Co}{55} &
         7.5 (-1)& 8.2 ( 0)& 3.3 ( 1)& 8.1 ( 1)& 1.9 (-1)& 1.5 ( 0)& 2.0 ( 1)\\
       \nucleus{\rm Fe}{56}($\nu_e,e^-$p)\nucleus{\rm Fe}{55} &
         5.4 ( 0)& 3.3 ( 1)& 1.0 ( 2)& 2.2 ( 2)& 2.2 ( 0)& 1.0 ( 1)& 7.3 ( 1)\\
       \nucleus{\rm Fe}{56}($\nu_e,e^-\alpha$)\nucleus{\rm Mn}{52} &
         6.1 (-2)& 9.8 (-1)& 4.9 ( 0)& 1.4 ( 1)& 9.9 (-3)& 1.2 (-1)& 2.5 ( 0)\\
       \nucleus{\rm Fe}{56}($\nu_e,e^-$)X &
         1.6 ( 1)& 7.4 ( 1)& 2.0 ( 2)& 4.1 ( 2)& 8.9 ( 0)& 2.9 ( 1)& 1.5 ( 2)\\
       \hline
       \nucleus{\rm Pb}{208}($\nu_e,e^-\gamma$)\nucleus{\rm Bi}{208} &
         4.7 ( 1)& 1.3 ( 2)& 2.5 ( 2)& 4.0 ( 2)& 3.5 ( 1)& 7.6 ( 1)& 2.2 ( 2)\\
       \nucleus{\rm Pb}{208}($\nu_e,e^-$n)\nucleus{\rm Bi}{207} &
         2.3 ( 2)& 9.9 ( 2)& 2.3 ( 3)& 4.0 ( 3)& 1.2 ( 2)& 4.2 ( 2)& 1.9 ( 3)\\
       \nucleus{\rm Pb}{208}($\nu_e,e^-$p)\nucleus{\rm Pb}{207} &
         1.8 (-2)& 1.1 (-1)& 3.3 (-1)& 6.9 (-1)& 7.2 (-3)& 3.3 (-2)& 2.3 (-1)\\
       \nucleus{\rm Pb}{208}($\nu_e,e^-\alpha$)\nucleus{\rm Tl}{204} &
         2.1 (-2)& 2.6 (-1)& 1.1 ( 0)& 3.0 ( 0)& 4.7 (-3)& 4.1 (-2)& 6.0 (-1)\\
       \nucleus{\rm Pb}{208}($\nu_e,e^-$)X &
         2.8 ( 2)& 1.1 ( 3)& 2.5 ( 3)& 4.5 ( 3)& 1.6 ( 2)& 4.9 ( 2)& 2.1 ( 3)\\
       \hline\hline
       \nucleus{\rm Fe}{56}($\overline{\nu}_e,e^+\gamma$)\nucleus{\rm Mn}{56} &
         3.4 ( 0)& 1.1 ( 1)& 2.2 ( 1)& 3.6 ( 1)& 2.3 ( 0)& 5.7 ( 0)& 1.9 ( 1)\\
       \nucleus{\rm Fe}{56}($\overline{\nu}_e,e^+$n)\nucleus{\rm Mn}{55} &
         5.0 (-1)& 4.5 ( 0)& 1.7 ( 1)& 4.2 ( 1)& 1.5 (-1)& 9.4 (-1)& 1.0 ( 1)\\
       \nucleus{\rm Fe}{56}($\overline{\nu}_e,e^+$p)\nucleus{\rm Cr}{55} &
         4.3 (-3)& 5.5 (-2)& 2.7 (-1)& 8.4 (-1)& 9.3 (-4)& 8.1 (-3)& 1.3 (-1)\\
       \nucleus{\rm Fe}{56}($\overline{\nu}_e,e^+\alpha$)\nucleus{\rm V}{52} &
         6.7 (-4)& 1.1 (-2)& 6.7 (-2)& 2.3 (-1)& 1.2 (-4)& 1.3 (-3)& 2.8 (-2)\\
       \nucleus{\rm Fe}{56}($\overline{\nu}_e,e^+$)X &
         3.9 ( 0)& 1.5 ( 1)& 3.9 ( 1)& 7.9 ( 1)& 2.4 ( 0)& 6.6 ( 0)& 2.9 ( 1)\\
       \hline
       \nucleus{\rm Pb}{208}($\overline{\nu}_e,e^+\gamma$)\nucleus{\rm Tl}{208} &
         5.8 (-1)& 3.0 ( 0)& 7.9 ( 0)& 1.5 ( 1)& 2.7 (-1)& 1.1 ( 0)& 6.1 ( 0)\\
       \nucleus{\rm Pb}{208}($\overline{\nu}_e,e^+$n)\nucleus{\rm Tl}{207} &
         4.9 (-1)& 3.8 ( 0)& 1.5 ( 1)& 3.9 ( 1)& 2.0 (-1)& 8.9 (-1)& 8.5 ( 0)\\
       \nucleus{\rm Pb}{208}($\overline{\nu}_e,e^+$p)\nucleus{\rm Hg}{207} &
         1.7 (-7)& 1.4 (-5)& 2.2 (-4)& 1.5 (-3)& 8.4 (-9)& 3.2 (-7)& 4.2 (-5)\\
       \nucleus{\rm Pb}{208}($\overline{\nu}_e,e^+\alpha$)\nucleus{\rm Au}{204} &
         4.3 (-7)& 4.0 (-5)& 6.5 (-4)& 4.4 (-3)& 2.1 (-8)& 8.1 (-7)& 1.2 (-4)\\
       \nucleus{\rm Pb}{208}($\overline{\nu}_e,e^+$)X &
         1.1 ( 0)& 6.8 ( 0)& 2.3 ( 1)& 5.4 ( 1)& 4.7 (-1)& 1.9 ( 0)& 1.5 ( 1)\\
       \hline\hline
    \end{tabular}
  \end{center}
  \caption{Total cross sections for charged current 
           neutrino scattering on nuclei for different neutrino energy
           spectra represented as Fermi-Dirac distributions. The 
           cross sections are given in units of $10^{-42} {\rm cm}^2$.}
\end{table}

\begin{table}
\begin{center}
\begin{tabular}{|r|l|l|l|l|} \hline \hline
 $E_\nu$ & 
 $^{56}$Fe($\nu_e,e^-$)X & 
 $^{56}$Fe($\nu,\nu'$)X & 
 $^{208}$Pb($\nu_e,e^-$)X &
 $^{208}$Pb($\nu,\nu'$)X    \\ \hline\hline
 10 & 6.61 (-1) &  1.91 (-1) & 9.34 ( 0) &  7.14 (-1) \\
 15 & 6.45 ( 0) &  2.19 ( 0) & 1.41 (+2) &  7.98 ( 0) \\
 20 & 2.93 (+1) &  6.90 ( 0) & 4.85 (+2) &  2.54 (+1) \\
 25 & 7.33 (+1) &  1.51 (+1) & 1.32 (+3) &  5.84 (+1) \\
 30 & 1.40 (+2) &  2.85 (+1) & 2.48 (+3) &  1.14 (+2) \\
 35 & 2.36 (+2) &  4.89 (+1) & 3.99 (+3) &  1.99 (+2) \\
 40 & 3.71 (+2) &  7.86 (+1) & 5.72 (+3) &  3.17 (+2) \\
 45 & 5.55 (+2) &  1.19 (+2) & 7.63 (+3) &  4.72 (+2) \\
 50 & 7.98 (+2) &  1.72 (+2) & 9.69 (+3) &  6.65 (+2) \\
 55 & 1.10 (+3) &  2.39 (+2) & 1.20 (+4) &  8.96 (+2) \\
 60 & 1.48 (+3) &  3.20 (+2) & 1.45 (+4) &  1.17 (+3) \\ 
 65 & 1.92 (+3) &  4.15 (+2) & 1.73 (+4) &  1.48 (+3) \\ 
 70 & 2.42 (+3) &  5.25 (+2) & 2.02 (+4) &  1.83 (+3) \\
 75 & 2.99 (+3) &  6.50 (+2) & 2.31 (+4) &  2.22 (+3) \\
 80 & 3.60 (+3) &  7.89 (+2) & 2.62 (+4) &  2.65 (+3) \\ 
 85 & 4.27 (+3) &  9.42 (+2) & 2.93 (+4) &  3.11 (+3) \\ 
 90 & 4.98 (+3) &  1.11 (+3) & 3.26 (+4) &  3.61 (+3) \\
 95 & 5.73 (+3) &  1.29 (+3) & 3.60 (+4) &  4.13 (+3) \\
100 & 6.52 (+3) &  1.49 (+3) & 3.96 (+4) &  4.69 (+3) \\ 
105 & 7.36 (+3) &  1.70 (+3) & 4.33 (+4) &  5.26 (+3) \\ 
110 & 8.24 (+3) &  1.92 (+3) & 4.71 (+4) &  5.86 (+3) \\
115 & 9.16 (+3) &  2.16 (+3) & 5.10 (+4) &  6.47 (+3) \\
120 & 1.01 (+4) &  2.41 (+3) & 5.50 (+4) &  7.09 (+3) \\ 
125 & 1.11 (+4) &  2.66 (+3) & 5.90 (+4) &  7.73 (+3) \\ 
130 & 1.21 (+4) &  2.92 (+3) & 6.31 (+4) &  8.37 (+3) \\
135 & 1.32 (+4) &  3.19 (+3) & 6.71 (+4) &  9.01 (+3) \\
140 & 1.42 (+4) &  3.46 (+3) & 7.12 (+4) &  9.66 (+3) \\ 
145 & 1.53 (+4) &  3.74 (+3) & 7.52 (+4) &  1.03 (+4) \\
150 & 1.64 (+4) &  4.01 (+3) & 7.91 (+4) &  1.09 (+4) \\ \hline\hline
\end{tabular}
\end{center}
\caption{Total $^{56}$Fe($\nu_e,e^-$)X, $^{56}$Fe($\nu,\nu'$)X,
         $^{208}$Pb($\nu_e,e^-$)X and $^{208}$Pb($\nu,\nu'$)X cross sections
         for selected neutrino energies $E_\nu$. 
         The cross sections are given in $10^{-42}$ cm$^2$, while
         the energies are in MeV. Exponents are given in parentheses}
\end{table}

\end{document}